\documentclass[conference]{IEEEtran}
\IEEEoverridecommandlockouts

\usepackage[ruled,linesnumbered]{algorithm2e}
\usepackage{algpseudocode}
\usepackage{multicol}

\usepackage[para,online,flushleft]{threeparttable}

\usepackage{pifont}
\usepackage{bbding} 

\def\BibTeX{{\rm B\kern-.05em{\sc i\kern-.025em b}\kern-.08em
    T\kern-.1667em\lower.7ex\hbox{E}\kern-.125emX}}
    
\usepackage{CJKutf8}  

\usepackage{xcolor}
\usepackage{pgfplots}
\usepackage{tikz}
\usepackage{comment}

\usepackage{filecontents}

\usepackage{fancyhdr} 
\usepackage{hyperref}

\usepackage{multirow}
\usepackage{cite}
\usepackage{amsmath,amssymb,amsfonts}

\usepackage{textcomp}
\usepackage{xcolor}

\usepackage{subfigure}
\usepackage{pgfplots}
\usepackage{filecontents}

\usepackage{graphicx,subfigure}

\definecolor{bblue}{HTML}{4F81BD}
\definecolor{rred}{HTML}{C0504D}
\definecolor{ggreen}{HTML}{9BBB59}
\definecolor{ppurple}{HTML}{9F4C7C}

\usepackage{listings}

\usepackage{blindtext}
\usepackage{etoolbox,xstring,mfirstuc,textcase}
\usepackage{booktabs}
\usepackage{pgfplots}

\usepackage{tabularx,booktabs,makecell,multirow, caption}
\usepackage{enumitem} 

\newcolumntype{L}{>{\arraybackslash}X}

\usepackage{xcolor}
\usepackage{tikz}
\usepackage{pgfplots}

\definecolor{findOptimalPartition}{HTML}{D7191C}
\definecolor{storeClusterComponent}{HTML}{FDAE61}
\definecolor{dbscan}{HTML}{ABDDA4}
\definecolor{constructCluster}{HTML}{2B83BA}

\usepackage[utf8]{inputenc}
\usepackage{listings}
\begin{document}
\title{Bridging BRC-20 to Ethereum} 


\author{\IEEEauthorblockN{Qin Wang, Guangsheng Yu, Shiping Chen}
\textit{CSIRO Data61,  Australia}
}

\maketitle

\begin{abstract}
In this paper, we design, implement, and (partially-) evaluate a lightweight bridge (as a type of middleware) to connect the Bitcoin and Ethereum networks that were heterogeneously uncontactable before. Inspired by the recently introduced Bitcoin Request Comment (BRC-20) standard, we leverage the flexibility of Bitcoin inscriptions by embedding editable operations within each satoshi and mapping them to programmable Ethereum smart contracts. A user can initialize his/her requests from the Bitcoin network, subsequently triggering corresponding actions on the Ethereum network. We validate the lightweight nature of our solution and its ability to facilitate secure and seamless interactions between two heterogeneous ecosystems.
\end{abstract}

\smallskip



\section{Introduction}

The emergence of Bitcoin revolutionized the field of financial technology by introducing a decentralized network for value transactions. Ethereum further advanced this concept by bringing smart contracts and decentralized applications (DApps). As of July 2023, the market capitalization of Bitcoin is approximately US\$584.97 billion, while Ethereum stands at around US\$223.32 billion (CoinMarketCap). These cryptocurrencies have not only created significant value themselves but have also paved the way for the development of numerous upper-layer tokens and DApps, which can reach market scales in the thousands. However, the structural differences between Bitcoin's UTXO model and Ethereum's account model have resulted in isolated ecosystems. Users are unable to freely transfer tokens between these heterogeneous blockchains and often rely on external intermediaries, such as centralized exchanges (CEX) or decentralized exchanges (DEX), which come with high costs and limitations. This lack of interoperability hinders the widespread adoption and evolution of these technologies, limiting their full potential.

Existing solutions have made efforts to facilitate interoperability among different blockchains. They often rely on various cryptographic techniques (e.g., zero-knowledge proofs~\cite{xie2022zkbridge} and hash-lock~\cite{wadhwahe}\cite{Herdius}), external hardware (e.g., TEE \cite{bentov2019tesseract}\cite{lan2021trustcross}) or reconstructing the entire system (e.g., Polkadot~\cite{wood2016polkadot}, Cosmos~\cite{kwon2019cosmos}). However, these approaches come with explicit limitations. Cryptographic approaches are computationally intensive and may introduce significant overhead. External hardware solutions like TEEs can be complex and difficult to implement. Reconstruction of the system requires extensive changes, bringing additional assumptions and complexities. As a result, current solutions suffer from various degrees of impracticability, which impede their wide adoption.

\smallskip
\noindent\textbf{Contributions.} To fill the gaps, we propose an innovative lightweight middleware protocol designed to bridge the gap between Bitcoin and Ethereum. The middleware takes advantage of BRC-20 \cite{brc20experiment}\cite{binancebrc}, an experimental standard for digital tokens on the Bitcoin network akin to Ethereum's ERC-20. Our idea is to interpret the BRC-20 operations inscribed on Bitcoin's blockchain (a.k.a., Bitcoin inscription~\cite{li2024bitcoin}) and reflect them on the Ethereum network, effectively extending Bitcoin's functionalities within Ethereum's EVM and enabling the possibility of integrating Bitcoin assets in DeFi applications~\cite{werner2022sok}\cite{jiang2023decentralized}. We approach the goal by completing the following steps:

\begin{itemize}
    \item \textit{We present a lightweight middleware, \textsc{MidasTouch}} (\textcolor{teal}{Sec.\ref{sec-protocol}}), designed to bridge the Bitcoin network and the Ethereum network. \textsc{MidasTouch} enables seamless communication (majorly from Bitcoin to Ethereum) between these networks, empowering users to interact with Ethereum smart contracts through Bitcoin inscriptions defined in the recent BRC-20 standard. 
    
    \item \textit{We have developed the preliminary version} of \textsc{MidasTouch} (\textcolor{teal}{Sec.\ref{sec-imple}}) to demonstrate its functionality. The prototype includes the implementation of functional methods on both the Bitcoin and Ethereum sides and the featured events in intermediate middleware, providing detailed insights into key operations.
    
    \item \textit{We conduct a partial evaluation} to assess the effectiveness and efficiency of \textsc{MidasTouch} (\textcolor{teal}{Sec.\ref{sec-evalua}}), focusing specifically on smart contract-related operations on the Ethereum testnet. Our evaluations are based on three key aspects: scalability and performance with varying committee sizes, gas usage for different contract functionalities, and the frequency of validator processing for requests. The results shed insights into the system's behavior under different scenarios, which align with people's intuitive expectations. Additionally, we have discussed the security aspects and potential limitations.
\end{itemize}

We emphasize two additional aspects in our designs:

\begin{itemize}
    \item \textit{U shape}. The workflow within our design takes the shape of a ``U": users initiate their requests by inputting inscriptions on the Bitcoin network. The action on inscriptions triggers a state transition within Ethereum. Eventually, the Ethereum contract concludes by furnishing a receipt to Bitcoin, serving as a record for the settlement process. 
    \item \textit{Lightweight}. The operation of the validator-maintained middleware does not inherently demand the involvement of supplementary participants. The validators responsible for upkeeping our middleware can either constitute the same group or form a subset of the Ethereum committee.
\end{itemize}

\noindent \textcolor{teal}{\ding{171}} We metaphorically refer to the task achieved by our middleware as \textsc{MidasTouch} (cf. our title), drawing inspiration from the tale in Greek mythology: \textit{everything King Midas touched turned to gold}, symbolizing a valuable connectivity effect.

\section{Before Construction}
\label{sec-preli}

\subsection{Building Blocks}

\noindent\textbf{BRC-20.} This Bitcoin-native standard \cite{brc20experiment}\cite{binancebrc} parallels the Ethereum ERC-20 token standard \cite{erc20} and signifies a significant shift within the Bitcoin ecosystem, particularly with the emergence of Bitcoin Ordinals. Bitcoin Ordinals revolutionize Bitcoin transactions by assigning an index to each Satoshi (the smallest unit of Bitcoin, 0.00000001) based on its mining order. These indices can be utilized for various purposes, such as unique identifiers or metadata, thereby unlocking new possibilities, including Non-Fungible Tokens (NFTs) \cite{erc721}\cite{wang2021non}. Once a Satoshi has been inscribed (TapScript, around 4MB), it can be utilized to create a BRC-20 token. In essence, the BRC-20 standard enables three primary operations: \textcolor{teal}{$\mathsf{deploy}$} (creation of a new token type),  \textcolor{teal}{$\mathsf{mint}$} (increasing the supply of the created token), and \textcolor{teal}{$\mathsf{transfer}$} (trading tokens). We provide a brief overview of each function below, with detailed information available in~\cite{wang2023understanding}. These functions collectively enable the creation of a simplified NFT implementation over the Bitcoin network,  albeit with some limitations in terms of extensibility.

\begin{lstlisting}[caption={ \textcolor{teal}{$\mathsf{Deploy}$-$\mathsf{Mint}$-$\mathsf{Transfer}$}\label{list-deploy} },basicstyle=\ttfamily\scriptsize]]
# Onchain Inscription (Identical items omitted)
"p" : "brc-20" # protocol name

"op": "deploy" # operation
"tick": "ordi" # token name
"max": "2100000" # total amount of token to be issued
"lim": "1000" # maximum amount of token minted each round

"op": "mint" # operation
"amt": "1000" # the amount of token being minted

"op": "transfer" # operation
"amt": "100" # the amount of token being transferred

# Off-chain update (mapping to operations)
"deploy": 
if state[tick] NOT exists:
    state[tick]={<inscription info>, 
                 "balances": {addr: balance_val}}

"mint": 
if state[tick] NOT exists OR 
                 "amt" > "lim" OR sum("amt") > "max":
    raise errors
else 
    account_state[tick]["balance"][minter] += amt

"transfer": 
if state[tick] NOT exists:
    raise errors
if state[tick]["balance"][sender] >= amt:
    account_state[tick]["balance"][sender] -= amt
    account_state[tick]["balance"][receiver] += amt
\end{lstlisting}

Due to the shortage of formal research on BRC-20, Rodarmor~\cite{bipord} was one of the first to introduce a scheme for assigning serial numbers to Bitcoin satoshis, and a comprehensive introduction to ordinal theory can be found in~\cite{ordinbook}. Additionally, Binance Research has published several pioneering reports~\cite{binancebrc}\cite{binance1}\cite{binance2} that explore the development of BRC-20. Bertucci~\cite{bertucci2023bitcoin} conducted an early analysis of transaction fees related to ordinal inscriptions.

\smallskip
\noindent\textbf{Smart contract.} A smart contract (SC) is a distinct form of contract where the agreement's terms are directly encoded into executable code. Operating as a self-contained \textit{white-box}, a smart contract guarantees the synchronization of input and output, effectively eliminating the reliance on trustworthy third-party intermediaries \cite{li2022smart}. Deployed primarily on blockchain platforms such as Ethereum, smart contracts are executed automatically once predetermined conditions encoded within the code are fulfilled.  The versatility of smart contracts enables automation across diverse domains, spanning from financial transactions \cite{werner2021sok} and governance systems \cite{kiayias2022sok} to decentralized organizations \cite{yu2023leveraging} and beyond. With their ability to enforce transparent and trustless transactions, smart contracts offer enhanced efficiency, security, and persistence.

\smallskip
\noindent\textbf{Ethereum token standards.} Tokens play a vital role in incentivizing users and developers within blockchain ecosystems. These tokens adhere to specific standards, which define the methods for creating, deploying, and issuing new tokens. Ethereum, with its robust smart contract capabilities, has established itself as a leader in token standards \cite{wang2022exploring}, driving versatile applications within its ecosystem. The ERC-20 fungible token standard \cite{erc20} has gained significant traction, leading to the proliferation of ICOs \cite{zetzsche2017ico} and a flourishing token market \cite{tapscott2017blockchain}. However, different blockchain ecosystems employ incompatible token standards. For instance, a token adhering to the BRC-20 standard on the Bitcoin network cannot be utilized on the Ethereum network. This limitation has motivated us to explore the construction of a potential connection between these disparate ecosystems.

\smallskip
\noindent\textbf{Bitcoin standards.} 
Bitcoin, operating as a standalone blockchain, lacks native token standards akin to Ethereum's ERC-20. However, proposals for tokenization methods exist within the Bitcoin Improvement Proposals (BIPs)~\cite{bips} (e.g., SegWit~\cite{bip141}\cite{bip84}, Taproot~\cite{bip341}\cite{bip342}). Establishing a Bitcoin standard is more rigorous than Ethereum, given Bitcoin's protocol's limited space for extension. BRC20, an external solution, predominantly handles complex functionalities off-chain, preserving minimal on-chain activity.

\subsection{Concurrent Solutions}

\noindent\textbf{Interoperability in blockchain.}
Polkadot \cite{wood2016polkadot} enables the interconnection of subnetworks (the relay chains) through the cross-chain message passing protocol (XCMP).  Within this context, a relay is a smart contract residing in a target blockchain that operates as a lightweight client of a source blockchain. Cosmos \cite{kwon2019cosmos} achieves cross-chain communications via the inter-blockchain communication protocol (IBC) \cite{ibc}. IBC is designed to consist of two major layers that establish secure connections for data transportation (\textit{a.k.a.} TAO) and define the way of packaging data (APP layer). However, these solutions are restricted to facilitating interoperability among blockchains within the same ecosystem. Cacti \cite{Cacti} is an integral part of the Hyperledger project. The scheme relies on a network of interoperable validators that validate cross-chain transactions and are entrusted with the task of signing them. To ensure the validity of transactions, a consensus among a quorum of validators is required for their successful signing. Hermes \cite{belchior2022hermes} is a middleware for blockchain interoperability that is built on the Open Digital Asset Protocol (ODAP) (recently merged into SATP \cite{satp}). The protocol draws inspiration from the two-phase commit protocol (2PC) \cite{bernstein1987concurrency} and goes beyond that by incorporating a flexible log storage API that provides multiple storage options, including local, cloud, and on-chain storage. CheaPay \cite{zhang2019cheapay} and Herdius \cite{Herdius} focus on a payment channel network that enables off-chain settlement of transactions between blockchains by utilizing the Hash Time-Lock Contract (HTLC) scheme to ensure atomic swaps of assets~\cite{xu2021game}\cite{narayanam2022atomic}. Tesseract \cite{bentov2019tesseract} is an exchange protocol that operates in real-time and utilizes trusted hardware as a reliable relay. It facilitates the tokenization of assets and enables the pegging of these assets to cryptocurrencies. Similar solutions also leverage TEEs \cite{lan2021trustcross}\cite{yin2022bool} to perform cross-chain transactions. Several studies put focus on the cross-chain bridges~\cite{notland2024sok}\cite{lee2023sok}. More classifications on interoperability can refer to \cite{belchior2021survey}\cite{wang2023exploring}.

\begin{table}[ht]
\caption{Mainstream interoperable blockchains}
\label{tab-interbc}
\centering
\begin{threeparttable}
\resizebox{\linewidth}{!}{
\begin{tabular}{l|cc|cc}
\multicolumn{1}{c|}{\rotatebox{0}{\textbf{\textit{\makecell{Projects}}}}} 
& \multicolumn{1}{c}{\rotatebox{0}{\textbf{\textit{\makecell{Communication}}}}} & 
\multicolumn{1}{c|}{\rotatebox{0}{\textbf{\textit{\makecell{Architecture}}}}} & 
\multicolumn{1}{c}{\rotatebox{0}{\textbf{\textit{\makecell{Witness}}}}} &
\multicolumn{1}{c}{\rotatebox{0}{\textbf{\textit{\makecell{Implementation}}}}} 
\\ 

\midrule

Polkadot  &  XCMP & Parachains &   Relay chain (SC) & Substrate \\
Cosmos  & IBC & Hybrid (TAO/APP)  & Relayers  &  Tendermint  \\
Hermes &  ODAP(-2PC)  & Gateway-based & Middleware & - \\ 
Hyperledger  & Trusted party & Hybrid &  Validators & Cactus \\ 
CheaPay  & Sidechain & Layer-2  & Hash-lock & -  \\
Herdius & Sidechain & Layer-2  & Hash-lock & -  \\
Tesseract  & Trusted hardware &  Hybrid &  TEE & (Exchange) \\ 
\bottomrule
\end{tabular}
}
\end{threeparttable}
\end{table}

\smallskip
\noindent\textbf{Selection of technical routes.} The presence of reliable witnesses is crucial for the successful implementation of a dependable interoperable protocol, especially in ensuring the \textit{all-or-nothing} settlement of digital assets, also known as atomic swaps. Existing solutions, as we have discovered through our investigation, rely on either trusted parties, such as relayers and validators, or automated algorithms/machines like smart contracts, middleware, TEEs, and hash-locks, to achieve reliable witnesses. However, relying on trusted parties poses a significant risk of compromise. Therefore, we have chosen the alternative route. Nonetheless, hash-locks are a common construct used in atomic cross-chain swap protocols (e.g., \cite{herlihy2018atomic}) which require strict network requirements, while TEE-based solutions tend to be complex. As a result, we have been motivated to develop a contract-based middleware that serves as an efficient bridge between Bitcoin and Ethereum, providing the desired functionality.

\subsection{Threat Model and Assumption}\label{sec-preli-aspt}

\noindent\textbf{Blockchain model.}  We assume the blockchains, both applicable to Bitcoin and Ethereum chains, consist of a group of nodes that operate the system. In our analysis, we simply consider a fraction of nodes that may behave arbitrarily, but the total number of these unfaithful nodes is below the security threshold of the consensus protocols (e.g., less than 50\% in PoS/PoW settings). The blockchain system adheres to the robust assumption as established by previous research~\cite{garay2015bitcoin}.

\begin{itemize}
\item \textit{Consistency.} Once an honest node commits transaction $tx_1$ before $tx_2$, no honest node will ever commit $tx_2$ before $tx_1$.

\item \textit{Liveness.} Once a valid transaction is submitted to an honest node, it will be eventually committed on-chain.
\end{itemize}

\smallskip
\noindent\textbf{Smart contract model.} By integrating the fundamental functionalities offered by underlying blockchain systems, we present a simplified abstraction of the key features to simulate a smart contract.

\begin{itemize}
\item \textit{Contract deployment.} The contract is deployed on-chain, establishing its presence within the blockchain network.

\item \textit{State update.} The state undergoes transitions triggered by input transactions, evolving as new transactions that are packed into the new proposed blocks.

\item \textit{State consensus.} Blockchain maintainers execute consensus to reach an agreement on the global view of the state, ensuring consistency among distributed nodes.

\item \textit{State query.} With a confirmed state, users can retrieve specific transactions and blocks at a given height for analysis or reference.
\end{itemize}

\smallskip
\noindent\textbf{Cryptography primitives.} We require conventional unforgeability for digital signatures (multi-signature  \cite{itakura1983public} included) and collision-resistant hash functions \cite{damgaard1989design}.

\smallskip
\noindent\textbf{General honesty assumption.} 
We make the assumption that the majority of group members in our described systems, whether they belong to the blockchain networks (such as Bitcoin and Ethereum) or the random validator committee, will faithfully adhere to the designated tasks. 

\section{Middleware Construction}
\label{sec-protocol}

\subsection{Technical Challenges}

\smallskip
\noindent\textbf{Challenge-I: Managing different data formats within heterogeneous blockchains is a non-trivial task.} The primary challenge lies in reconciling the stateless UTXO-based model of Bitcoin with Ethereum's stateful, account-based approach. To address this, we propose leveraging the BRC-20 standard as a middleware to establish a lightweight bridge.

In our implementation, BRC-20 utilizes inscriptions to record the state transitions of UTXO states. These inscriptions serve as verifiable proofs and are used to trigger smart contract functions/events. We incorporate a series of operation indicators within the inscriptions and provide corresponding events in the smart contract. This allows users to initiate transactions on the Bitcoin network, which in turn triggers changes in the inscriptions and subsequent state transitions in the smart contracts on the Ethereum network.

\smallskip
\noindent\textbf{Challenge-II: Determining which side should bear the deduction of fees can give rise to debatable arguments.} State-of-the-art cross-chain solutions often overlook the costs involved in exchanges, which is a crucial aspect that users are concerned about.

In our approach, the actions are initiated from the Bitcoin side, and the actual transactions are triggered on this network, leading to corresponding state transitions on the Ethereum side. Consequently, users on the Bitcoin side are responsible for bearing the associated exchange fees, which are separate from (equiv. in addition to) the basic transaction fees incurred during the consensus procedures.

\smallskip
\noindent\textbf{Challenge-III: Implementing cross-chain transactions may pose significant complexity.} Existing schemes often rely on a series of complex cryptographic operations (e.g., \cite{bentov2019tesseract}) or the reconstruction of intricate systems (e.g., \cite{wood2016polkadot}\cite{kwon2019cosmos}). Unfortunately, this level of complexity renders the system impractical for widespread adoption and use.

Rather than introducing complex dependencies, our approach focuses on establishing a lightweight middleware that seamlessly bridges actions initiated on the Bitcoin side with state transitions on the Ethereum side. Our implementation leverages the native editable field in Bitcoin, as defined by the BRC-20 standard, and programmable functions written in smart contracts. \textsc{MidasTouch} can work harmoniously with both blockchains, ensuring smooth interoperability without the need for additional intricate dependencies.

\subsection{Warm-up Construction}

\smallskip
\noindent\textbf{Roles.}  
The protocol includes four roles: \textit{transaction originator}, \textit{contract owner}, \textit{validator}, and \textit{operator}. 

 \begin{itemize}
    \item  \textit{Transaction originator}  \textcolor{teal}{(Bitcoin)}. The Bitcoin transaction originators are the users initiating transactions on the Bitcoin network. Their main role is to inscribe the transaction with specific information regarding Ethereum contract interactions, such as contract address and operation data. This inscription is embedded within Bitcoin transactions and is scanned by validators.

    \item  \textit{Contract owner} \textcolor{teal}{(Ethereum)}. The Ethereum contract owners control the smart contracts on the Ethereum network, which the middleware protocol interacts with. They define the contract operations that can be invoked through inscriptions on the Bitcoin network. Furthermore, they monitor the state updates broadcast by validators.
  
    \item  \textit{Validator} \textcolor{teal}{(middleware)}. The validators are responsible for the accurate execution of the middleware protocol. Their duties include registering themselves on the list, validating transactions from the Bitcoin network, and managing the update of Ethereum contract states. They also participate in consensus processes. Notably, validators have to deposit an amount of money in the contract.


    \item  \textit{Operator} \textcolor{teal}{(middleware)}. The operators are responsible for setting up and maintaining the middleware protocol. They set the system parameters, such as the size of the validator committee, the block intervals for consensus and state updates, the rules for multi-signature validation, and other security features. They also take care of system upgrades.
    
\end{itemize}

Note that in a typical case, middleware \textit{validators}, \textit{operators}, and the Ethereum \textit{contract owner} can be played by the same group of users, wherein settings established by the middleware \textit{operator} are commonly predefined in the genesis or a designated checkpoint block.

\begin{figure}[!htb]
    \centering
	\includegraphics[width=0.99\linewidth]{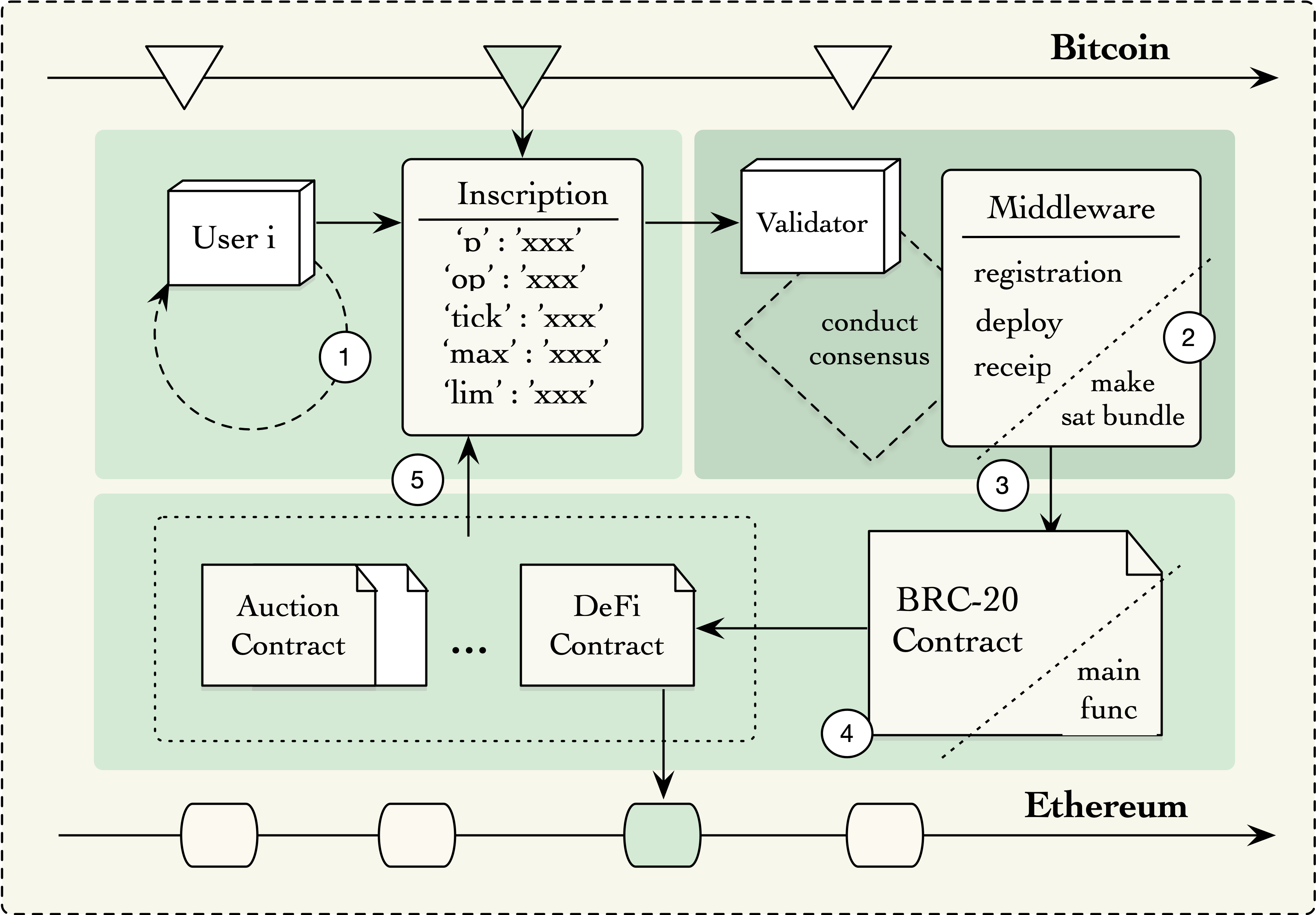}
	\caption{System overview}\label{fig_mid}
\end{figure}

\smallskip
\noindent\textbf{System overview} (Fig.\ref{fig_mid}). 
For the initial setup, the contract developer deploys a domain-specific smart contract on Ethereum. In this protocol, each token does not have its own smart contract. Smart contracts are organized based on functionality, where each functionality (such as the auction function in DeFi) has its own smart contract. Each contract records the state information of all tokens that use that particular functionality.

Then, \ding{172} any Bitcoin user who is keen to become a validator of the middleware layer will deposit a certain amount of ETH to this smart contract with a valid Ethereum address that is bounded with his Bitcoin address. Until enough validators get registered (while the validator committee size is pre-defined and can be ranged from one to multiple), the system initialization is completed. The committee will take charge of all interaction with the related smart contracts, on behalf of the Bitcoin transaction originators who thus do not need to own any Ethereum addresses. A user sends an inscribed satoshi that claims the validity of any functions (equiv. operations). This sat should contain a script (TapScript). The functions, formatted as \textsl{func\_signature} (as in the example), need to match the unified interfaces defined in corresponding smart contracts. \ding{173} The committee consistently monitors the Bitcoin network and periodically collects the inscribed satoshi (acts as an explorer), sorts them as per the timestamp, and constructs a sat bundle. \ding{174} For every increment of $\varepsilon$ block heights in the Bitcoin network, the committee engages in a consensus process related to the sat bundle. Following this, it invokes the respective Ethereum network's smart contracts using the contract addresses corresponding to each sat bundle. \ding{175} The protocol employs a multi-signature approach to update the state in the smart contracts. Auditing is processed to ensure that the penalty is properly applied (by slashing the deposit) to any misbehaviors. Meanwhile, the gas fee awarded to validators will be deducted from the satoshi bundle with a certain percentage, e.g., 5\%. Note that the gas fee will be calculated for each individual satoshi. \ding{176}  Finally, the committee gathers the emitted events from the invoked contracts and broadcasts the post-operation inscriptions on the Bitcoin network. These broadcasts act as receipts for the executed bundle, signaling the completion of the originated inscription requests.

\begin{lstlisting}[caption=Example format,basicstyle=\ttfamily\scriptsize]]
"operation": name # matched to interfaces in contracts
"smart contract address": xxx # Ethereum
"deposit": xxx ETH # Ethereum token
...
\end{lstlisting}

\subsection{Detailed Construction}

\begin{itemize}

\item \textbf{\textit{Validator registration.}} The \textsc{MidasTouch} protocol initiates with defining core parameters and proceeds to the validator registration phase. Validators are required to register and deposit a specified amount of ETH into a designated deposit contract. The registration is inscribed on the Bitcoin network, after which the newly registered validator is added to the validator set. The size and requirements of the validation committee can vary significantly based on the desired level of system security. For instance, a system requiring high security might necessitate a large committee and more intricate consensus mechanisms. Conversely, a system prioritizing efficiency over security might operate with a smaller committee or even a single validator.

\item \textbf{\textit{Inscription-contract interactions.}} Once the validator committee is established, the middleware protocol begins managing transactions from the Bitcoin network. For each output in every transaction of the newly obtained Bitcoin block, it searches for potential inscriptions. Valid inscriptions are added to the \textit{inscription bundle set} $\mathbb{B}$, and the corresponding contract addresses are accumulated into the contracts set in terms of different functionalities.

\item \textbf{\textit{State update.}} The consensus process (if the validator committee size reaches the lower bound for consensus) and state update occur at predetermined block intervals. During this consensus process, the inscriptions bundle is sorted based on the timestamp, and validators reach a consensus on the legitimate inscriptions. The system then fetches the latest state for each contract in the set from the Ethereum network, which commonly include a balance record for each unique Bitcoin address associated with various tokens as an entry point for handling necessary operations upon the token amounts.

\item \textbf{\textit{Multi-signature validation.}} The protocol processes each inscription within the bundle, subtracting the gas fee from each and distributing it among validators based on their respective Bitcoin addresses. If the operation within the inscription proved valid in the Bitcoin network, it is executed with multi-signature validation, leading to the state and address balance updates in the Ethereum network. The degree of validation and the consensus mechanism used for this process can be adjusted according to the security requirements of the system.

\item \textbf{\textit{Inscription publication.}} After processing all contracts, validators republish the outcomes of operations as inscriptions back to the Bitcoin network. The block index is incremented, indicating the protocol's readiness to manage the next Bitcoin block.

\end{itemize}

\section{Implementation}
\label{sec-imple}

\subsection{Basic Operations} \label{subsec-operation}
We provide three concrete instances to clarify the proposed \textsc{MidasTouch} protocol. The operations cover $\mathsf{registration}$, token $\mathsf{deploy}$, and $\mathsf{receipt}$ production.

\begin{itemize}
  \item $\mathsf{Registration}$. The operation includes details such as the \textit{protocol name}, \textit{operation} (registration), \textit{signature}, \textit{token name}, \textit{deposit amount}, and \textit{Ethereum address}. Following the inscription, an update occurs on Ethereum. If the Ethereum address does not exist in the validator set, it is added with the associated Bitcoin address and balance information. The connection between Bitcoin and Ethereum addresses forms the backbone of the middleware.

\begin{lstlisting}[caption={$\mathsf{Registration}$},basicstyle=\ttfamily\scriptsize]]
# Bitcoin Inscription
"p" : "middleware", # protocol name
"op": "registration", # operation
"op_signature": "registration(...) return (...)", 
          # operation signature
"tick": "eth", # token name of deposit
"max": "32", # total amount of deposit
"c_addr": <eth_addr> # Ethereum contract address of
                     # the designated deposit contract
"eth_addr": <eth_addr>
           # Ethereum address used by a validator 
           # in registration

# Ethereum update
# Contract_address: c_addr
function registration(...) return (...) {
    ...
    if validator[eth_addr] NOT exists:
        validator[eth_addr]={"btc_addr": <btc_addr>, 
                  "balances": <balance_val>} }
\end{lstlisting}

  \item $\mathsf{Deploy}$. The operation represents issuing a new token on the Ethereum network. The Bitcoin inscription includes information including the \textit{protocol name} (in this case, the BRC-20 token standard), \textit{operation} (deploy), \textit{operation signature}, \textit{token name}, \textit{total token supply}, and the \textit{upper bound for tokens} to be minted in each round. The Ethereum network responds to this inscription by creating a new corresponding token state if it does not exist. This state contains the inscription information and an initial balance state for the token addresses.

\begin{lstlisting}[caption={$\mathsf{Deploy}$},basicstyle=\ttfamily\scriptsize]]
# Bitcoin Inscription
"p" : "brc-20", # protocol name
"op": "deploy", # operation
"op_signature": "deploy(...) return (...)", # signature
"tick": "ordi", # token name
"max": "2100000", # total amount of tokens to be issued
"lim": "1000" # maximum amount of token being minted
"c_addr": <eth_addr> # Ethereum contract address of
                     # the targeted contract

# Ethereum update
# Contract_address: c_addr
function deploy(...) return (...) {
    if state[tick] NOT exists:
        state[tick]={<inscription info>, "balances": 
                    {addr: balance_val}} }
\end{lstlisting}

  \item $\mathsf{Receipt}$. The operation represents the closure of an Ethereum event cycle and its report back to the Bitcoin network to guarantee the finalization of the originated inscription requests such as the above $\mathsf{Deploy}$ inscription. On the Ethereum side, after a function defined in the smart contract is executed, events are emitted. These events are captured by validators, who then publish an inscription on the Bitcoin network, signifying an operation of $\mathsf{receipt}$. This inscription includes the \textit{protocol name} and a collection of \textit{events}, each corresponding to an \textit{inscription ID} and carrying information about the operation's \textit{execution results} (e.g., true/false, values). Unlike other algorithms, there is no ``op\_signature'' in this case, as this algorithm is simply forwarding the Ethereum events back to the Bitcoin network without executing a particular operation itself. With this operation, only those inscription requests which are included in an $\mathsf{receipt}$ will be committed as a success and ready to be further used.

\begin{lstlisting}[caption={$\mathsf{Receipt}$},basicstyle=\ttfamily\scriptsize]]
# Ethereum events emitted
# Contract_address: <eth_addr>
function <func_signaure> {
    ...
    emit();
}

# Bitcoin Inscription
"p" : "middleware", # protocol name
"op": "receipt", # operation
"events": {inscription_id: (<t/f>, <return_value>)}
\end{lstlisting}
\end{itemize}

\subsection{Algorithms} \label{subsec-algorithm}

We implement and present the major workflow of \textsc{MidasTouch} (cf. \textcolor{black}{Algorithm~\ref{algo-1}}), featured with a multi-party validator committee. Notations are listed in Table~\ref{tab-notation}.

\smallskip
\noindent\textbf{Become committee members.} Becoming an eligible committee member requires \textcolor{teal}{$\mathsf{CommiteeRegistration}$} to be involved. The \textcolor{teal}{$\mathsf{CommiteeRegistration}$} function is responsible for registering validators who hold both a Bitcoin and Ethereum address and have deposited a specific amount of ETH into a contract. This registration is then inscribed onto the Bitcoin network, and the newly registered validator is added to the validator set $\mathbb{V}$. The function also confirms the successful registration of these validators and ensures the completeness and correctness of the information provided during the registration process.

\smallskip
\noindent\textbf{Action on Bitcoin.} The primary function involved in the actions on the Bitcoin network is \textcolor{teal}{$\mathsf{HandleInscription}$}. This function scans each transaction output from the new Bitcoin block for potential inscriptions. Valid inscriptions are appended to the transaction bundle $\mathbb{B}$, and the corresponding contract addresses are accumulated into the contracts set $\mathbb{C}$. Additionally, \textcolor{teal}{$\mathsf{HandleInscription}$} is also responsible for broadcasting post-operation inscriptions, which serve as receipts for the executed transactions, indicating their completion. This dual functionality ensures that all potential inscriptions are evaluated for validity and corresponding receipts are issued, keeping the system secure and transparent.

\smallskip
\noindent\textbf{Action on Ethereum.} The main function involved in the actions on the Ethereum network is \textcolor{teal}{$\mathsf{UpdateEVMState}$}. This function is responsible for retrieving the most recent state $\mathbb{S}$ for every contract $c$ within the set $\mathbb{C}$ from the Ethereum network. For contracts related to BRC-20 or those possessing similar token-managing functionalities, it additionally retrieves a balance record $\Psi$ for each distinct Bitcoin address associated with various tokens.  During the consensus process that occurs every $\varepsilon$ block, this function processes each inscription within the bundle $\mathbb{B}$, distributing the gas fee $g$ among validators based on their respective Bitcoin addresses and updating the state $\mathbb{S}$ and address balance $\Psi$ via a multi-signature validation process. Furthermore, \textcolor{teal}{$\mathsf{UpdateEVMState}$} oversees the gathering of emitted events from the invoked contracts on Ethereum. These events are later broadcasted on the Bitcoin network as part of the receipt operations, marking the completion of inscriptions.

\begin{table}[ht]
\caption{Notations}
\label{tab-notation}
\centering
\begin{threeparttable}
\resizebox{1\linewidth}{!}{
\begin{tabular}{c|c|c}
\multicolumn{1}{c|}{\rotatebox{0}{\textbf{\textit{\makecell{Symbol}}}}} 
& \multicolumn{1}{c|}{\rotatebox{0}{\textbf{\textit{\makecell{Meaning}}}}} & 
\multicolumn{1}{c}{\rotatebox{0}{\textbf{\textit{\makecell{Scope}}}}} 
\\ 

\midrule

$\mathbb{V}$ & validator set, instantiated by $v_i$  & \textsc{MidasTouch}  \\
$\mathbb{S}$ & state set, instantiated by $s_i$  &  Ethereum \\
$\mathbb{C}$ & smart contact set, instantiated by $c_{\textrm{addr}}$  &  Ethereum \\
$\mathbb{B}$  &  inscription bundle set & Bitcoin \\
$c_{\textrm{addr}}$ & contract address/identifier & Ethereum \\
$\Lambda$ & receipt set &  Bitcoin/Ethereum  \\
$\Psi$ & address balance &  Bitcoin \\
$H$ &  block height/index & Bitcoin/Ethereum\\
$p$ & penalty rate &  Bitcoin \\
$g$ & gas fee & Ethereum  \\
$\varepsilon$ & constant value & Bitcoin/Ethereum  \\
$ins$ & short for inscriptions & Bitcoin \\
$tx$ & short for transactions & Bitcoin/Ethereum \\
 $\phi$  &  validator mapping topology  & Bitcoin/Ethereum\\
\bottomrule
\end{tabular}
}
\end{threeparttable}
\end{table}

\begin{algorithm*}[!hbt]
\scriptsize
\caption{The \textsc{MidasTouch} Protocol}\label{algo-1}
\SetAlgoLined
\begin{multicols}{2}[\raggedcolumns]
\textbf{Initialization} 

$H: \text{{block index}}$, $\mathbb{V}: \text{{validators}} \gets \emptyset$, $\mathbb{S}: \text{{state}} \gets \emptyset$, $\Psi: \text{{Bitcoin's balance}} \gets \emptyset$, $p: \text{{penalty rate}}, g: \text{{gas fee rate}}$, $\mathbb{B}: \text{Bundle}, \mathbb{C}:  \text{Contracts} \gets [], []$\;

\SetKwFunction{MainFunc}{\textcolor{teal}{$\mathsf{MainFunc}$}}
\SetKwFunction{CommitteeRegistration}{\textcolor{teal}{$\mathsf{CommitteeRegistration}$}}
\SetKwFunction{HandleInscription}{\textcolor{teal}{$\mathsf{HandleInscription}$}}
\SetKwFunction{UpdateEVMState}{\textcolor{teal}{$\mathsf{UpdateEVMState}$}}

\SetKwProg{firstproc}{Procedure}{}{}
\firstproc{\MainFunc{}}{
\CommitteeRegistration{}\;

\While{True}{
  $\text{{Block}} \gets \mathbb{V}.\text{{GetBitcoinBlock}}(H)$\;

  \ForEach{$\text{tx}: \text{{transaction}}$ in $\text{Block}.\text{{tx}}$}{
    \ForEach{$o: \text{{output}}$ in $\text{tx}.\text{{outputs}}$}{
      $ins: \text{{inscription}} \gets \mathbb{V}.\text{{ParseInscription}}(o)$\;
      \If{$ins \neq \text{{None}}$}{
        $\mathbb{B}.\text{{append}}(\HandleInscription(ins))$\;
        $\mathbb{C}.\text{{append}}(\HandleInscription(ins)[\text{c\_addr}])$;
      }
    }
  } 

  \If{$H \mod \varepsilon == 0$}{
  
  $\mathbb{B} \gets \mathbb{V}.\text{{Sort}}(\mathbb{B}, \text{key=timestamp}).\text{{ConsensusProcess}}(\mathbb{B}, p)$\;
  \ForEach{\text{c\_addr} in \text{$\mathbb{C}$} \text{parallelly}}  
  {
    $\mathbb{S}, \Psi \gets \mathbb{V}.\text{GetStateInfo(\text{c\_addr})},
    \mathbb{V}.\text{GetBalance(\text{c\_addr})}$;
  
    $\mathbb{S}, \Psi, \text{True/False} \gets \UpdateEVMState(\mathbb{B}, \mathbb{S}, \Psi, g)$;

    (\text{True/False}) events of \textit{c\_addr} are cached in the receipt set $\Lambda$\;
  }
  $\mathbb{V}.\text{PublishInscription}(\Lambda)$, \text{each event referred to its own $ins$};\

$\mathbb{B}: \text{Bundle}, \mathbb{C}:  \text{Contracts} \gets [], []$\;
}
  
  $H \gets H + 1$\;
} 
}\textbf{end}

\SetKwProg{secondproc}{Procedure}{}{}
\secondproc{\CommitteeRegistration{}}{
  \While{$\mathbb{V}.\text{{count}}() < \text{min\_committee\_size}$}{
    $\phi: (map[\text{{BTC\_addr}}]\Leftrightarrow \text{ETH\_addr}) \gets \mathbb{V}.\text{{WaitForRegistration}}()$\;
    $ins: \text{{inscription}} \gets \mathbb{V}.\text{{ParseInscription}}(\phi.\text{BTC\_addr.tx})$\;
    \If{$\mathbb{V}.\text{{GetDeposit}}(\phi.\text{{ETH\_addr}}) \geq k (\text{ETH})$ \text{in a dedicated deposit contract}  and $ins$ \text{is for registration and finalized}}{
      $\mathbb{V}.\text{{append}}(\phi)$\;
    }
  }
} 

\SetKwProg{thirdproc}{Procedure}{}{}
\thirdproc{\UpdateEVMState($\mathbb{B}, \mathbb{S}, \Psi, g$)}{
    $\text{{True/False}} \gets [], []\;$
    
  \ForEach{$ins: \text{{inscription}}$ in $\mathbb{B}$}{
    $\text{gas\_fee} \gets g \times ins[\text{value}]$\;
    $ins[\text{value}] \gets ins[\text{value}] - \text{gas\_fee}$\;
    
    \ForEach{$\phi: \text{validator in }\mathbb{V}$}
    {$\Psi[ins[\text{c\_addr}]][\phi.\text{BTC\_addr}][ins[\text{tick}]] \gets \Psi[ins[\text{c\_addr}]][\phi.\text{BTC\_addr}][ins[\text{tick}]] + \text{gas\_fee}$\;
    }
    
    \If{$ins[\text{operation}] \in \mathbb{C}[ins[\text{c\_addr}]].\text{{GetInterfaces}}()$}{
        \text{Invoke $ins[\text{operation}]$ with multi-sig validation};
    }
    \text{True.append}($ins$) \textbf{if }\textit{succeeded} \textbf{else} \text{False.append}($ins$)\;
  }
  \Return{$\mathbb{S}, \Psi$, \text{True/False}}\;
} 

\SetKwProg{fourthproc}{Procedure}{}{}
\fourthproc{\HandleInscription{$ins: \text{{inscription}}$}}{
  $\text{{op}} \gets ins[\text{operation}]$, $\text{{value}} \gets ins[\text{values}]$ \;
  $\text{{c\_addr}} \gets ins[\text{c\_addr}]$, $\text{{others}} \gets ins[<others>]$\;
  
  \Return{\text{op}, \text{c\_addr}, \text{value}} 
   } \textbf{end}
\end{multicols}
\end{algorithm*}

\subsection{Use Case}

To illustrate the real-world operation of the \textsc{MidasTouch}, we provide the following case. We have three participants named Alice, Bob, and Carol, all of whom are actively engaged in the network and aspiring to become validators for the system. To be eligible, they utilize the \textcolor{teal}{$\mathsf{CommiteeRegistration}$} function. Each participant provides their Bitcoin and Ethereum addresses and deposits a specified amount of ETH into a designated contract $c$. This transaction is recorded or inscribed onto the Bitcoin network, and subsequently, Alice, Bob, and Carol are added to the validator set $\mathbb{V}$.

Then, we introduce Dave, an end-user who intends to execute a transaction on the Bitcoin network. Dave creates an inscription in the transaction output, which is included when the new block is mined. At this point, the \textcolor{teal}{$\mathsf{HandleInscription}$} function comes into play. It scans each transaction output from the newly mined Bitcoin block, validates Dave's inscription, and appends it to the bundle $\mathbb{B}$. The corresponding contract address is also added to the contracts set $\mathbb{C}$.

On the Ethereum network, the \textcolor{teal}{$\mathsf{UpdateEVMState}$} function is triggered every $\varepsilon$ block. This function retrieves the latest state $s_i$, where $s_i\in  \mathbb{S}$, for each contract $c$ within the set $\mathbb{C}$, including the contract to which Dave's inscription was added. In the case where the contract is associated with BRC-20 or similar token-managing functionalities, the function also fetches the balance record $\Psi$ for each unique Bitcoin address linked to various tokens, including Dave's address.

During each $\varepsilon$ block, Alice, Bob, and Carol, as validators, process each inscription within the bundle $\mathbb{B}$. They distribute the gas fee $g$ among themselves based on their respective Bitcoin addresses. Through a collaborative multi-signature validation process, they update the state $s'_i$ ($s'_i\in\mathbb{S}$) and address balances $\Psi$. This completes the entire workflow of the proposed middleware protocol, ensuring a consistent state is maintained across the Bitcoin and Ethereum networks.

\section{Evaluation and Analysis}
\label{sec-evalua}

\subsection{Performance Analysis}

\begin{figure}[!hbt]
\centering
\includegraphics[width=9cm]{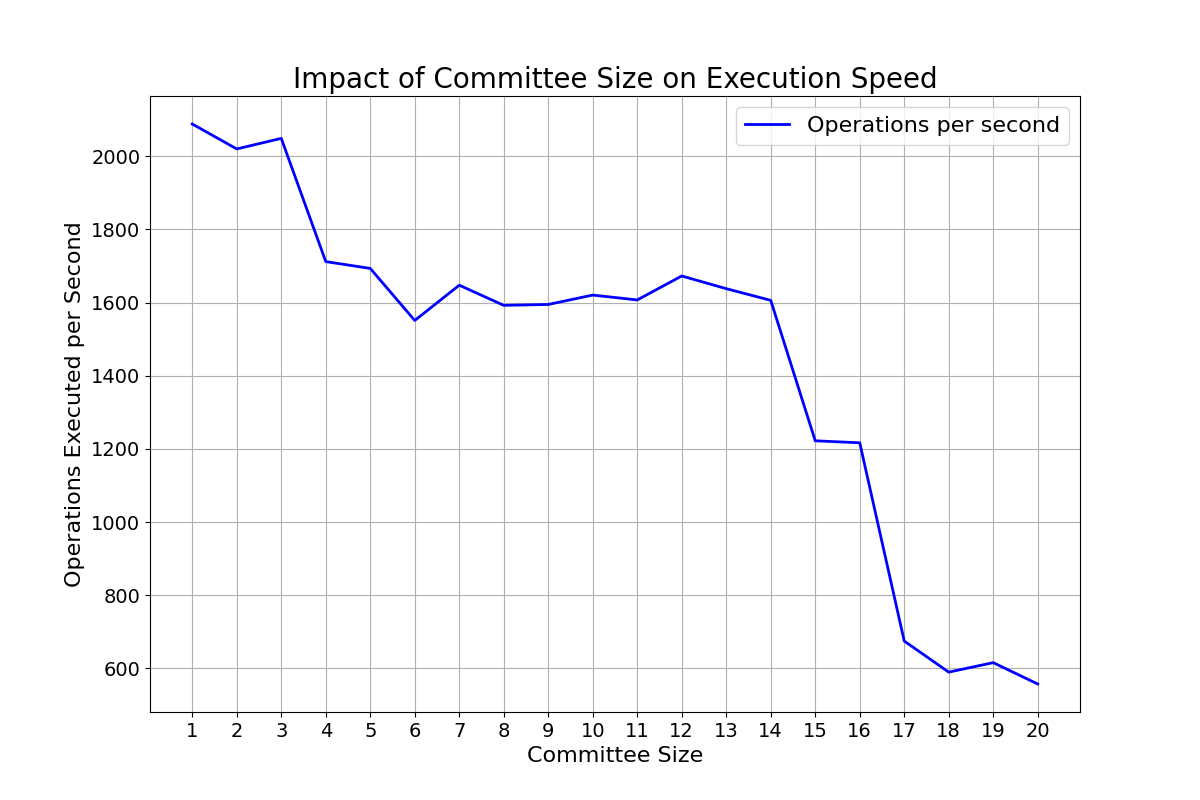}
\caption{Evaluation on scalability}
\label{fig:scalability}
\end{figure}

\noindent\textbf{Scalability.}
Fig.\ref{fig:scalability} presents a detailed visualization of the impact of the (validator-) committee size on the execution speed of our proposed \textsc{MidasTouch} protocol. The x-axis represents the size of the committee, which ranges from 1 to 20 members\footnote{A validator committee with a size smaller than 4 is regarded as a central entity with no consensus process being done.}. The y-axis illustrates the number of operations that can be executed per second. We utilize the well-known Practical Byzantine Fault Tolerance (PBFT) \cite{CastroL99} protpcol for our committee and factor in the transaction processing capabilities of both Bitcoin and Ethereum networks. Specifically, we consider Bitcoin's Lightning Network which can process up to the order of 10,000 transactions per second~\cite{cryptoeprint:2022/1614}, and Ethereum 2.0 where the Casper-PoS~\cite{buterin2019casper}\cite{buterin2020combining} and sharding technology are enabled, capable of handling 64 times higher transactions per second than the single-sharded Ethereum when the projected Phase-1 is activated. Given these parameters, the operational speed of \textsc{MidasTouch} cannot surpass the minimum throughput of these two networks.

As depicted in Fig.\ref{fig:scalability}, the execution speed corresponds to Ethereum 2.0's average throughput until a committee size of 4, the minimum requirement for consensus in our configuration. After this point, the speed decreases non-linearly with an increase in committee size due to the quadratic time complexity ($O(n^2)$) of the PBFT, where $n$ represents the number of nodes.
This illustrates that the choice of committee size presents a balancing act between decentralization and performance. While larger committees yield increased decentralization, they compromise on operational speed.

\smallskip
\noindent\textbf{Gas consumption.} We evaluate the additional amount of gas that any inscription needs to pay for validators in terms of the different functionalities of smart contracts running on the Ethereum network. Note that the overhead of sending an inscription is small and can be negligible when compared with the execution of smart contracts due to which an incentive is required to be applied. We consider the gas consumption in typical contracts regarding each functionality:
\begin{itemize}
    \item FT (fungible tokens~\cite{erc20}). The simplest type of smart contract typically involves just the transfer of tokens from one address to another.

    \item NFT (non-fungible tokens~\cite{wang2021non}): NFT contracts can be complex due to the involvement of metadata handling, uniqueness verification, or royalty payment mechanisms.

    \item Stablecoin~\cite{clark2019sok}: Generally simple as well, but some additional complexity for pegging the value to an asset.

    \item Insurance: It can get complicated depending on the terms of the insurance policy and the type of risks it covers.

    \item Loan~\cite{black2019atomic}: Loan contracts can be complicated. They usually require mechanisms to handle interest calculation, risk assessment, and loan recovery.

    \item Auction: Auction contracts need to manage bids from multiple participants, which adds complexity.

    \item DAO: These are the most complex types of contracts involving governance, voting mechanisms, fund management, or interacting with many other types of contracts.

\end{itemize}

\vspace{-0.8em}
\begin{figure}[!hbt]
\centering
\includegraphics[width=1\linewidth]{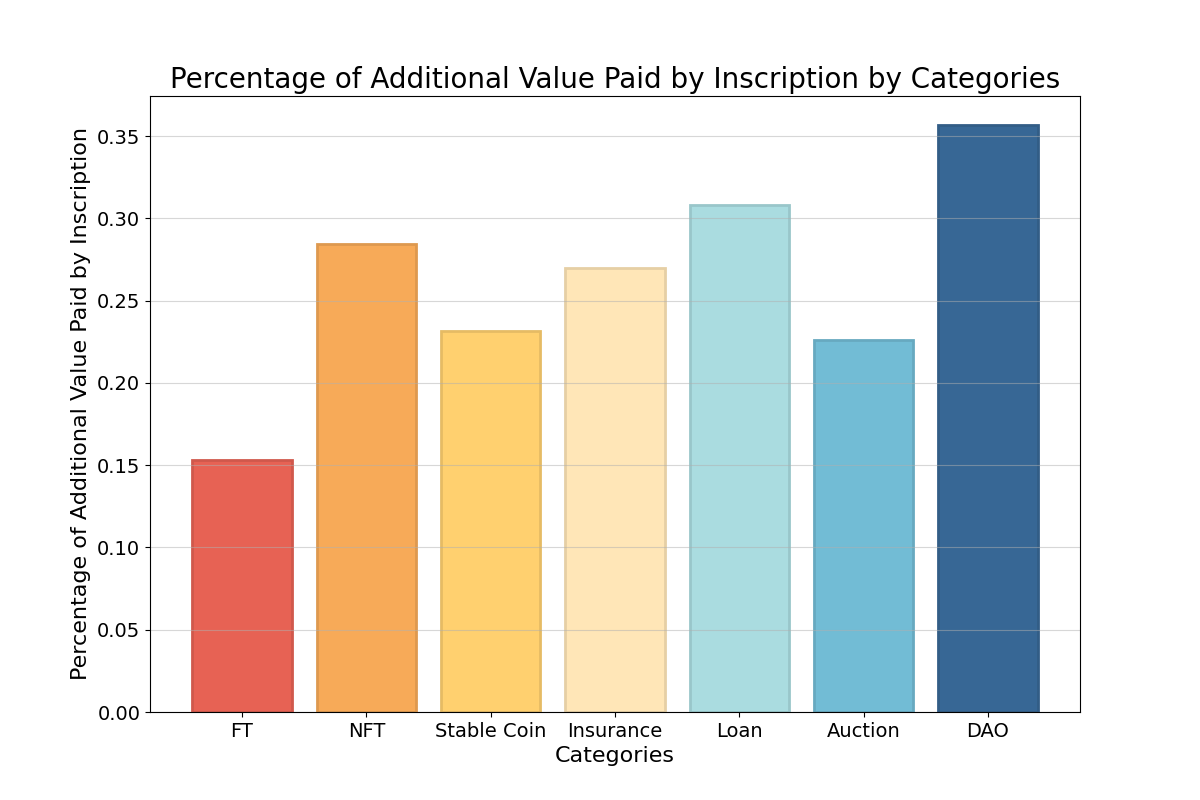}
\centering
\caption{Evaluation on gas consumption for finalization}
\label{fig:gasused}
\end{figure}

Specifically, the percentage of additional value paid by inscription across various categories of smart contracts is presented in Fig.\ref{fig:gasused}, using representative types as examples. FT requires the least additional value, reflecting their relatively straightforward functionality of merely transferring tokens. In contrast, DAOs~\cite{wang2022empirical}, with their intricacies involving governance, voting mechanisms, and fund management, demand the highest percentage. NFTs, Loans, and Auctions lie in the middle ground. NFT contracts' complexity arises from handling metadata and verifying uniqueness, while Loan contracts necessitate mechanisms for calculating interest, assessing risk, and recovering loans. Auction contracts, owing to their need to manage multiple participants' bids, also require a substantial additional value. It is noteworthy that the additional value percentages are influenced by the inherent complexity and functionality of the respective smart contracts. This insight underscores the necessity for efficient management of gas consumption in order to maximize the overall system efficiency.

\begin{figure}[!hbt]
\centering
\includegraphics[width=0.9\linewidth]{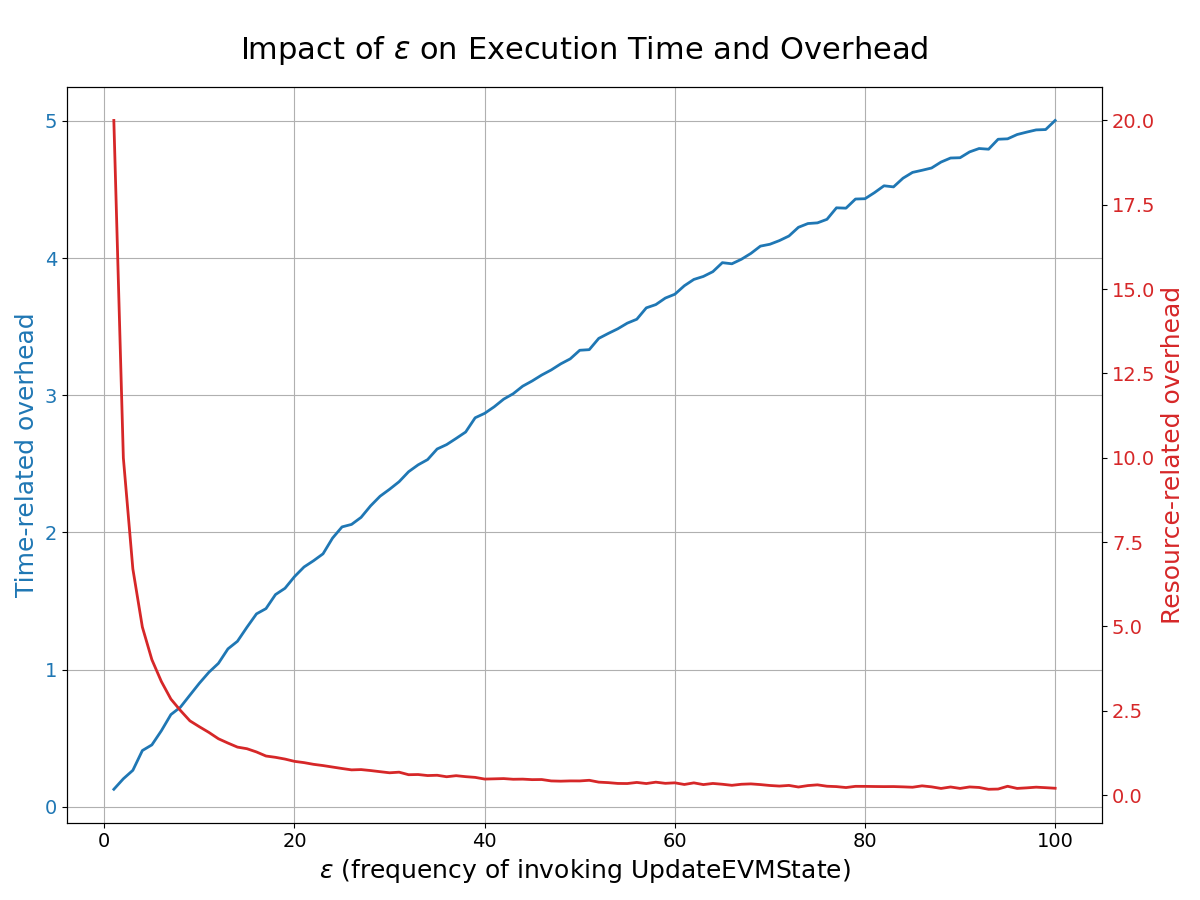}
\caption{Evaluation on different numbers of inter blocks}
\label{fig:varepsilon}
\end{figure}

 \smallskip
\noindent\textbf{Frequency of checking.} 
We further explore the influence of the parameter $\varepsilon$, which dictates the frequency of invoking the \textcolor{teal}{$\mathsf{UpdateEVMState}$} operation in terms of Bitcoin block heights, on the efficiency of the \textsc{MidasTouch} protocol. Fig.\ref{fig:varepsilon} demonstrates two distinctive aspects of system overhead: time-related overhead, and resource-related overhead, which are associated with the execution time and the computational resources required, respectively.

When $\varepsilon=1$, the validator committee is obligated to scrutinize every Bitcoin block, extract the inscriptions from transactions, assemble them into a bundle, arrange them by timestamp, and finally update the corresponding Ethereum smart contracts. As an alternative, the system can postpone the update of Ethereum's smart contract state until every $\varepsilon$ Bitcoin block heights have been processed, amassing a substantial number of sorted inscriptions in the bundle during this interval.

In addition, both the time-related and resource-related overheads are affected by the choice of $\varepsilon$. Specifically, as $\varepsilon$ increases, the time-related overhead decreases gradually, demonstrating that accumulating more transactions before updating the EVM state can save execution time. However, this comes at the cost of an increase in resource-related overhead, likely due to the need for storing and sorting a larger number of inscriptions. The ideal value of $\varepsilon$ would therefore be a trade-off between these two factors, balancing the need for quick execution with the capacity of the system's available resources.

\vspace{-0.8em}
\vspace{-0.8em}

\subsection{Primary Security Analysis}

\noindent\textbf{Safety (settlement).}
In our context, the appropriate definition of safety extends beyond conventional distributed systems definitions that primarily focus on state consistency. Here, safety implies that a request is fully executed on both sides and returns the correct value. To achieve this, we primarily ensure settlement: a request should be treated as an indivisible unit that can reach the final state on both sides. This guarantees that the protocol remains in a consistent state and prevents the fraudulent creation of additional values or unwarranted destruction of legitimately existing values within the Ethereum network. The complete lifecycle in our protocol is marked by two Bitcoin transactions: the first transaction (where an inscription request is included) serves as a trigger, initiating a series of events; and the second transaction acts as a receipt, indicating the successful completion of all associated events. 

Firstly, the unidirectional invoking nature of the \textsc{MidasTouch} protocol guarantees that Bitcoin users can successfully complete token transfers from the origin address to the designated address. This indicates that the operations recorded in the inscription can be invoked. Given our assumption that the majority of validators are honest, these operations will faithfully transit through our channel. Upon reaching the smart contract, the operations are processed on the Ethereum chain based on the endorsement of the majority of validators through multi-signature verification. Secondly, after the execution of operations, receipts are issued and broadcasted on the Bitcoin network to guarantee the finalization of the originated inscription requests included in the executed bundle, providing an additional layer of security and transparency.

Furthermore, the \textsc{MidasTouch} protocol does not possess the ability to externally deduct digital assets from either side. Transactions on Bitcoin are initiated by users, while the legal invocation of a contract requires the majority of validators. Any misbehavior will be rejected through internal consensus procedures. Thus, the addition of the receipt operation ensures the protocol's safety and settlement, offering conclusive evidence of successful transaction executions.

\smallskip
\noindent\textbf{Liveness.}
The liveness property of the system ensures that it remains continuously available without the risk of sudden shutdown. In the context of our protocol, this property signifies that any actions performed within the Bitcoin network, such as transactions or inscriptions, will eventually be reflected in the Ethereum network. This property relies on the correct functionality of the majority of validators, which we assume to be guaranteed (our assumptions in Sec.\ref{sec-preli-aspt}).

\smallskip
\noindent\textbf{Fairness.}
The property ensures equality among valid transactions, prohibiting any discrimination. This indicates any Bitcoin transaction conforming to protocol rules and having paid the necessary gas fee will be finally reflected in Ethereum without bias. Fairness can be achieved through the settlement of operation processing on both sides.

\section{Further Discussions}

\noindent\textbf{Utility in ``one-direction''.} Our bridge is designed to facilitate unidirectional interactions, enabling users to invoke actions and trigger state transitions from the Bitcoin network to the Ethereum network. It is important to note that our solution does not support bi-directional functionality, meaning that users must initiate the workflow from the Bitcoin side and follow the established pathway to trigger events within the Ethereum smart contract. While the unidirectional nature of our bridge may impose certain constraints on its scope of application and usage, it is an inherent limitation resulting from the distinct data formats utilized by Bitcoin and Ethereum. The use of UTXOs in Bitcoin ensures a reliable transaction ordering mechanism but, by its nature, restricts the support for other features such as general state transitions in contracts. However, despite this limitation, we have successfully established a lightweight directional channel at a minimal cost, offering valuable assistance to Bitcoin users seeking to interact with the Ethereum network.

\smallskip
\noindent\textbf{Evaluation on ``one-side''.} Our evaluation primarily focuses on the smart contract functionality and performance within the Ethereum testnet. The limitation arises from our consideration of costs, particularly due to the unaffordability of conducting batch transactions on Bitcoin's network, which lacks a dedicated testnet. In this initial version, we have implemented all the functional events on both the Bitcoin and Ethereum sides, enabling us to maximize our evaluation of the potential performance and associated costs. However, we acknowledge that there is ample room for further optimization. We encourage industry teams with an interest in this topic to invest more resources into conducting comprehensive evaluations.

\smallskip
\noindent\textbf{Extension, ``NOT" comparison.} Even though we propose a middleware to bridge the Bitcoin network and Ethereum, our primary emphasis is not on cross-chain functionality, but rather on leveraging Ethereum as a tool to enhance BRC20. As a result, the protocol has been intentionally crafted to address the specific requirements of the BRC-20 scenario.

\smallskip
\noindent\textbf{Faithfulness of validators.} It is well recognized that even permissioned blockchain systems are not completely immune to the trustworthiness of validators, regardless of the committee size. Concerns may arise among regular users regarding the potential compromise of validators, which could pose a threat to the stability of the middleware. To mitigate such risks, we recommend that each middleware validator deposits a substantial amount of tokens (e.g., 32 ETH \cite{ethstaking}) into the protocol. This ensures that validators have significant stakes in the network, reducing the likelihood of malicious behavior. This will provide users with a higher level of confidence when transferring larger amounts of tokens through the middleware. Additionally, increasing the committee size by enabling dynamic formation can significantly enhance the robustness and decentralization of the system, moving it closer to a \textit{permissionless} model~\cite{pass2017rethinking}. However, it's important to acknowledge that some degree of centralization might persist~\cite{yuzhe2023how}, but steps can be taken to mitigate this tendency.

\section{Conclusion}
\label{sec-conclusion}
Current Bitcoin and Ethereum are isolated due to their heterogeneous chain structure. In this work, we propose a lightweight one-way middleware, named \textsc{MidasTouch}, to bridge the Bitcoin and Ethereum networks. We employ the notion of the newly proposed BRC-20 standard to incorporate a range of operations into each satoshi and associate them with specific events within Ethereum smart contracts. We implement a prototype of  \textsc{MidasTouch} and evaluate the performance from the Ethereum side. Evaluation results demonstrate practicability and efficiency. To our knowledge, this is the first attempt to expand the capabilities of BRC-20.

{\footnotesize \bibliographystyle{IEEEtran}
\bibliography{bib}}






\end{document}